# CO$_2$-induced Rejuvenation in Polyetherimide: a New Key to Understand the Brittle-to-Ductile Transition in Mechanical Behavior of Nanocellular Polymers.

*Félix Lizalde-Arroyo, Frederik Van Loock, Victoria Bernardo, Miguel Ángel Rodríguez-Pérez, Judith Martín-de León\**

F. Lizalde-Arroyo, V. Bernardo, M.A. Rodríguez-Pérez, J. Martín-de León
CellMat Laboratory, Campus Miguel Delibes, Faculty of Science. Condensed Matter Physics Department, University of Valladolid, Valladolid, Spain
BioEcoUVa Research Institute on Bioeconomy, University of Valladolid, Valladolid, Spain
Email: judit.martin.leon@uva.es

F. Van Loock
Processing and Performance of Materials, Department of Mechanical Engineering, TU Eindhoven, Eindhoven, the Netherlands

Funding: Financial support from FPI grant PRE2022-101933 (Félix Lizalde-Arroyo) from the Spanish State Investigation Agency and MCIN/AEI/10.13039/501100011033 and European Social Fund Plus (ESF+) are gratefully acknowledged. Financial assistance from Ministerio de Ciencia, Innovación y Universidades (MCIU) (Spain) (PID2021-127108OB-I00, TED2021-130965B-I00, and PDC2022-133391-I00), Regional Government of Castilla y León and the EU-FEDER program (CLU-2019-04) are gratefully acknowledged. This work was supported by the Regional Government of Castilla y León (Junta de Castilla y León), and by the Ministry of Science and Innovation MICIN and the European Union Next GenerationEU / PRTR. (C17.I1).

**Keywords:**

**Abstract:**
Nanocellular polyetherimide exhibits significant improvements in mechanical properties like toughness and impact resistance, commonly associated with the presence of nanoporosity. However, this work demonstrates these enhancements, often measured directly after processing, cannot be fully explained solely by the cellular structure but also originate from a modification of the polymer matrix induced by the $CO_2$-saturation process. Through a systematic study involving thermal treatments and saturation-desorption processes without foaming, it is shown that $CO_2$ exposure, even in the absence of pore formation, induces an apparent rejuvenation of the polymer, as evidenced by a reduction in the yield stress, which persists after complete $CO_2$ desorption and in the absence of residual gas during mechanical testing. Therefore, the observed ductile response is not associated with the presence of $CO_2$ during deformation, but with a permanent modification of the polymer matrix induced by prior gas exposure. This structural state can be thermally reversed by activating the β-relaxation of the polymer. For nanocellular polymers, the presence of residual gas within the matrix during foaming restricts thermal relaxation and helps preserve the $CO_2$-modified state. As a result, the mechanical response of the solid phase reflects the intrinsic properties of the saturated polymer and the architecture imposed by the cellular structure. This work demonstrates for the first time that $CO_2$ saturation can permanently alter the mechanical state of a high-$T_g$ amorphous polymer, providing a new framework to interpret the brittle-to-ductile transition in nanocellular PEI.

# 1. Introduction

Nanocellular polymers are advanced porous materials combining low density, robust mechanical performance, thermal insulation, and sometimes optical transparency, making them promising for applications such as aerospace, automotive, railway, and advanced electronics.[1–4] These materials are defined by pore/cell sizes ($\phi$) smaller than 1 μm, which induce confinement effects in both the solid and gaseous phases. For instance, reducing pore size to a few hundred nanometers triggers the Knudsen effect, significantly lowering the thermal conductivity of the confined gas in porous.[5–7] In addition, it was also proved that this confinement significantly affects the solid phase of the material. Several studies confirm that, when the thickness of the pore walls drops below a threshold (which depends on the polymer) there is an appreciable immobilization of polymer chains, accompanied by an increase in glass transition temperature ($T_g$) and a reduction in the free volume between chains.[4,8,9] These structural changes at the molecular level were correlated with improved mechanical properties at the macroscopic scale. Moreover, when the pore size is further reduced below the critical threshold of 50 nm, additional phenomena arise. In this regime, a transition occurs in the light scattering mechanism, from Mie to Rayleigh scattering, enabling the fabrication of optically transparent porous materials.[10] This subclass of material in which the cell size is extremely low has been termed ultra-nanocellular polymer, which exhibits not only emergent properties such as transparency but also enhanced effects derived from confinement in both phases. These materials represent a step forward in the field of porous polymers, due to the combination of low density with advanced thermal, optical, and mechanical performance.[11]

However, the fabrication of such structures, especially when using high-performance polymers (HPPs) and aiming for the extremely small cell sizes required, remains a significant challenge.[1,12] The high chain rigidity, elevated value of $T_g$, and limited segmental mobility of these materials hinder both nucleation and cell growth at the nanometric scale.[13] In this context, gas dissolution foaming has proven to be the most effective and versatile technique for producing nanocellular polymers with precisely controlled morphology. This method, which involves saturating the polymer matrix with $CO_2$ under controlled saturation pressure ($P_{sat}$) and temperature ($T_{sat}$) conditions, followed by rapid depressurization and thermal foaming, enables the formation of nano- and even ultra-nanocellular structures.[14] This is made possible thanks to the dual role of the gas, as a blowing agent and as a temporary plasticizer, which transiently reduces the effective glass transition temperature ($T_{g,eff}$) and enhances chain mobility during cell expansion.[15]

In recent years, despite the complexity associated with the fabrication of nanocellular polymers from HPPs, several studies have demonstrated the successful application of this method to produce nanopores in polyphenilsulfone (PPSU) or polyetherimide (PEI), among others, achieving structures with pore sizes between 20 - 80 nm and density reductions up to ranges between 30 and 50% for PPSU, and 40 and 60% for PEI.[16–18] Moreover, major improvements in mechanical performance were proven in these materials in comparison to the solid precursor, especially in impact resistance, with increases of up to close to 65%, and elongation at break ($\varepsilon_b$) in a uniaxial tension test, where the best result for a nanocellular PEI reached a fracture strain of 1.34 compared to 1.10 for the solid, often attributed directly to the presence of the nanoscale morphology.[3,4] Nevertheless, despite significant advances in structural design and the understanding of the physical properties of these materials, there remains a considerable gap in the interpretation of the origin of their enhanced mechanical behavior.[2]

However, several hypotheses were proposed in the literature. From a molecular perspective, the reduction of pore size to the nanometric scale has been shown to induce a three-dimensional confinement of polymer chains within the solid phase of the cellular polymer. This confinement limits segmental mobility and increases intermolecular interactions, which was correlated with an increase in the $T_g$ and the local stiffness of the polymer matrix. Such effects were experimentally observed in nanocellular poly(methyl-methacrylate) (PMMA), where local measurements using AM-FM AFM (Amplitude Modulation - Frequency Modulation Atomic Force Microscopy) revealed a rise in the Young's modulus at the nanometer scale as pore size decreased, along with lower values of the loss tangent (*tan δ*) values, further indicating reduced segmental mobility in the confined polymer regions.[19] Additionally, recent investigations combining Attenuated Total Reflectance Fourier Transform Infrared Spectroscopy (ATR-FTIR), Raman spectroscopy, and Differential Scanning Calorimetry (DSC) confirmed that molecular confinement within pore walls thinner than 100 nm for PMMA leads to a measurable reduction in chain polarizability and a shift toward higher vibrational frequencies, consistent with shorter effective bond lengths and increased local stiffness. These molecular-level changes correlate with elevated $T_g$ and denser chain packing in the solid phase.[8] Similar trends were reported in PEI-based nanocellular structures, where a rise in $T_g$ was observed as the pore size enters into the nanoscale, reinforcing the role of confinement in the solid phase.[4] Yet, at the macroscopic level, the improvements observed in properties such as toughness, energy absorption, or elongation at break in nanocellular polymers, particularly in materials like PEI, have been attributed almost exclusively to the presence of the nanostructure itself.[4] For instance, it has been proposed that improvements in impact resistance and fracture may be related to enhanced ductility of the polymer struts between the nano-scale pores.[20] Furthermore, during macroscopic deformation, the stretching, collapse, or rearrangement of the pores may promote a densification processes, thereby reducing damage localization and contributing to the apparent increase in ductility and impact resistance.[2,4] This behavior was also reported in nanocellular fluorinated polyimide (FPI) foams, where pore sizes below 50 nm resulted in a remarkable combination of optical transparency, mechanical flexibility, and structural integrity. These improvements were explicitly attributed to a reduction in local stress concentrations facilitated by the nanoscale morphology, and were supported by both experimental data and finite element simulations.[3] Moreover, similar effects were observed in commodity polymers such as PMMA, although in this case, no direct causal link to the nanostructure was established. In solid-state foamed PMMA with pore sizes ranging from 20 to 84 nm and relative densities between 0.37 and 0.5, a notable increase in fracture toughness was reported in comparison to microcellular polymers (~1 μm and similar densities to the nanocellular materials produced). Through uniaxial compression and single-edge notch bending (SENB) tests, it was shown that while the relative Young's modulus and yield strength remain unaffected by pore size, the relative fracture toughness increases as the pore size decreases.[9]

An implicit assumption in most of these studies is that the improved ductility arises from the nanocellular architecture itself. However, in all these materials, the polymer matrix is necessarily exposed to high concentrations of $CO_2$ prior to foaming. Whether this exposure can permanently modify the polymer matrix, even after complete gas desorption, has not been addressed so far.

Moreover, achieving nanocellular morphologies typically requires the application of elevated temperatures during foaming, often above or below $T_g$, and it is known that heating affects the polymer chains' relaxation behavior. However, these thermal processes typically lead to accelerated non-equilibrium relaxation processes (known as physical aging) which tend to increase the yield stress, contrary to what is observed in foamed materials.[21,22]

For this reason, this work proposes that the saturation step, after cell nucleation, may reset prior thermomechanical history of the polymer at the molecular level. To investigate this hypothesis, a systematic study is proposed to compare the mechanical behavior of solid PEI with samples subjected exclusively to $CO_2$ saturation and full desorption, and to thermal treatments below and above $T_g$. In addition, $CO_2$-saturated samples subsequently exposed to thermal treatment were also investigated. The aim is to identify whether additional mechanisms are responsible for the transition in mechanical response typically observed in nanocellular polymers, beyond the presence of the cellular structure itself.

Based on these considerations, we hypothesize that $CO_2$ saturation drives the polymer into a rejuvenated structural state that is retained after full desorption, and that this state largely affects the mechanical response of the solid phase in nanocellular PEI.

## 2. Results and Discussion
### 2.1. Evaluation of Mechanical Behavior in Nanocellular PEI

In this study, PEI-based materials were fabricated using the gas dissolution foaming technique, as detailed in *Experimental Section* and **Table 1**. Three types of samples were analyzed: untreated (as-received) solid PEI, an opaque nanocellular PEI with an average cell size of approximately 780 nm, and a transparent ultra-nanocellular PEI with an average cell size of approximately 35 nm. The two porous samples had similar relative densities (~0.82). The results obtained from uniaxial tensile tests on the three types of materials, conducted according to the methodology outlined in *Experimental Section,* are presented in **Figure 1a**. Although the maximum ductility values do not differ drastically among samples, a shift in average failure behavior is observed as pore size decreases: from an average fracture strain equal to 0.40 for the untreated solid PEI, to 0.56 for the nanocellular PEI, and 0.57 for the ultra-nanocellular PEI. Additionally, in cellular materials, fracture occurs at comparatively higher and more consistent strain values, whereas in untreated solid PEI the strain at break is lower and exhibits a larger standard deviation, as reported in **Figure 1a** where each value is shown together with its standard deviation in the color of the corresponding curve. This suggests a change in the deformation and fracture behavior of the nano- and ultra-nanocellular samples compared to the untreated solid PEI. These results follow a trend consistent with previous findings in the literature.[4]

**Figure 1b** provides an even clearer insight. Impact tests (see details of the procedure in *Experimental section*) reveal that the ultra-nanocellular PEI (relative density ~0.82) absorbs more energy (3.1 J/mm) than the solid (1.9 J/mm) and exhibits a distinct more ductile failure mode as seen in the pictures included in **Figure 1b**. Although the absolute values obtained in this test may differ from those reported in other studies, these discrepancies are primarily due to differences in the equipment and test method used, distinct from those employed in previous works such as Miller et al.[4] Nevertheless, the most significant observation is the transformation in the failure mechanism: while the untreated solid tends to fracture in a brittle and abrupt manner, the ultra-nanocellular PEI shows a more progressive and controlled response, with an enhanced ability to dissipate energy during impact. This property could be especially advantageous in applications requiring protection against dynamic loads, such as aerospace components, electronics, or transportation structures exposed to vibrations or impacts.[23] These findings are consistent with those reported by Miller and co-workers (**Figure 1c**), where a similar behavior was observed in nanocellular PEI materials produced using the same gas dissolution foaming method.[4] At a relative density of 0.90, the energy absorbed by the ultra-nanocellular polymer (20.5 J/mm) was approximately 65% higher than that of the untreated solid (12.5 J/mm), and approximately six times higher than that of the equivalent microcellular polymer (3.2 J/mm). For densities of 0.83 and 0.75 and nanocellular structures, improvements of 31% and 10% over the solid were reported, respectively, and more than 190% and 50% compared to their microcellular counterparts. Once again, the nanocellular structure samples led to a clear transition in fracture mode from brittle to ductile as seen in the pictures shown in **Figure 1c**. To date, this improvement has been attributed exclusively to nanostructured morphology.

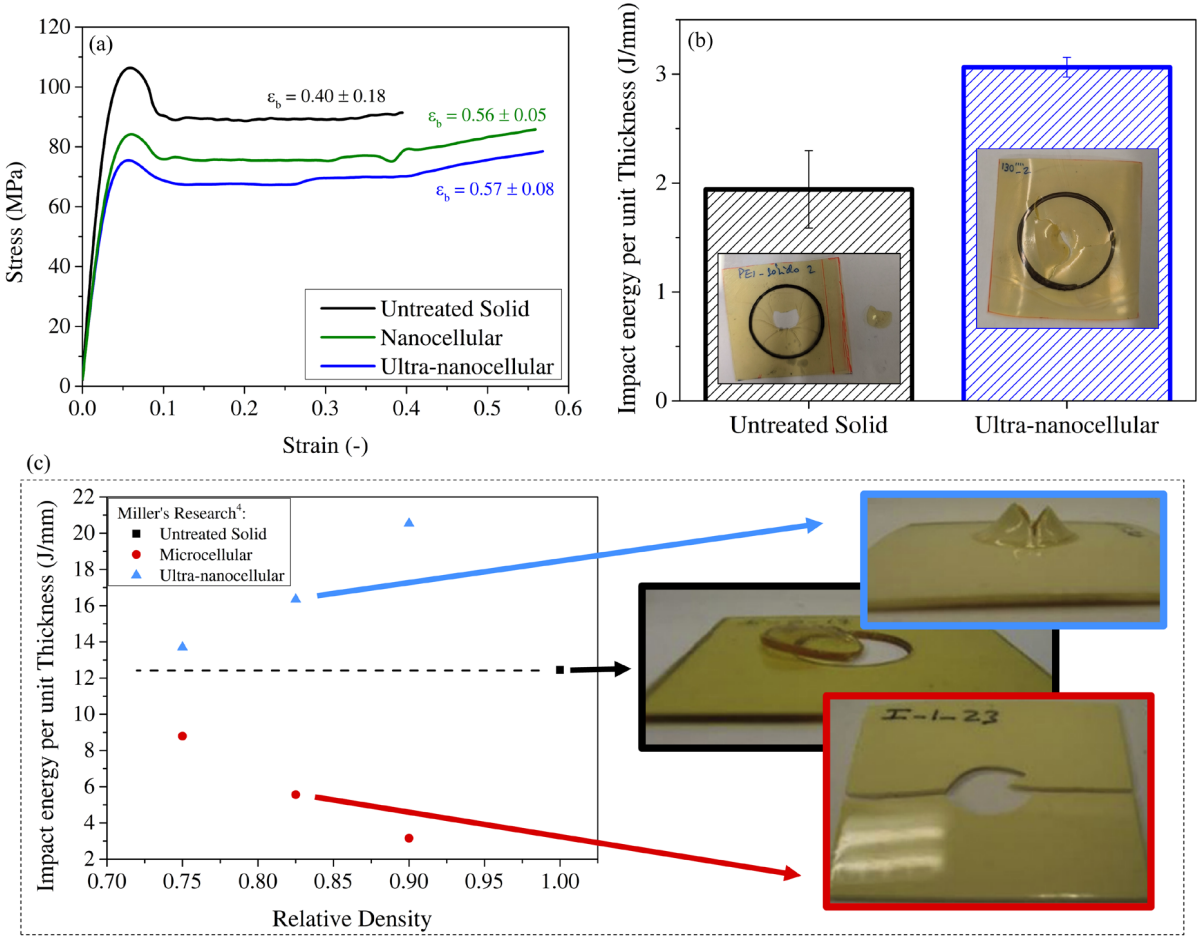

**Figure 1**. (a) Representative uniaxial tensile stress-strain data for untreated solid PEI, nanocellular PEI, and ultra-nanocellular PEI when deformed in uniaxial tension. The strain at break ($\varepsilon_b$) is reported for each case, following the same color plotted for the curve, including

the corresponding standard deviation. (b) Energy absorbed during falling dart impact tests, accompanied by images of the fractured samples that illustrate the change in failure mode across different materials. (c) Impact results reported by Miller et al. (2011), showing a similar transition in mechanical behavior. Original images from reference 4 of their fractured samples are included to highlight morphological similarities in the fracture surfaces shown in (b).[4]

However, from a molecular perspective, this interpretation could remain incomplete. There is no compelling reason why the mere presence of pores, even at the nanoscale, should inherently lead to a sustained increase in ductility in a high-$T_g$ amorphous polymer. These highlights need to consider other factors in the fabrication process that could alter the internal structure of the polymer. In the case of gas dissolution foaming, two critical steps may induce molecular reconfiguration, as shown in the boxed area of **Figure 2**: (c) $CO_2$ saturation, where the gas acts not only as a blowing agent but also as a temporary plasticizer that enhances chain mobility; and (d) thermal foaming, in which the material is exposed to elevated temperatures, below its reference value of $T_g$ (when no $CO_2$ is present) but above the $T_{g,eff}$ of the plasticized, $CO_2$-saturated polymer.[18] Both phenomena, acting separately or together, may leave a lasting imprint on the solid polymer matrix. To isolate and better understand these effects, the following sections provide a systematic analysis of how heat and $CO_2$ saturation, independently and in the absence of cellular structure, affect the properties of solid PEI. **Figure 2** also summarizes this experimental approach, illustrating the different processes applied to the samples, from untreated polymer (a) and isolated thermal treatments (b) to $CO_2$ saturation (c), followed by either gas desorption (e) or thermal foaming (d), and subsequent mechanical testing of the specimens subjected to these processes (f). This schematic highlights the steps in which molecular reconfiguration of the solid phase can occur and how they are connected within the overall study design. So, **Figure 2**, in conjunction with **Table 1** from the *Experimental Section*, provides a comprehensive overview of the processes and experimental conditions applied in this study.

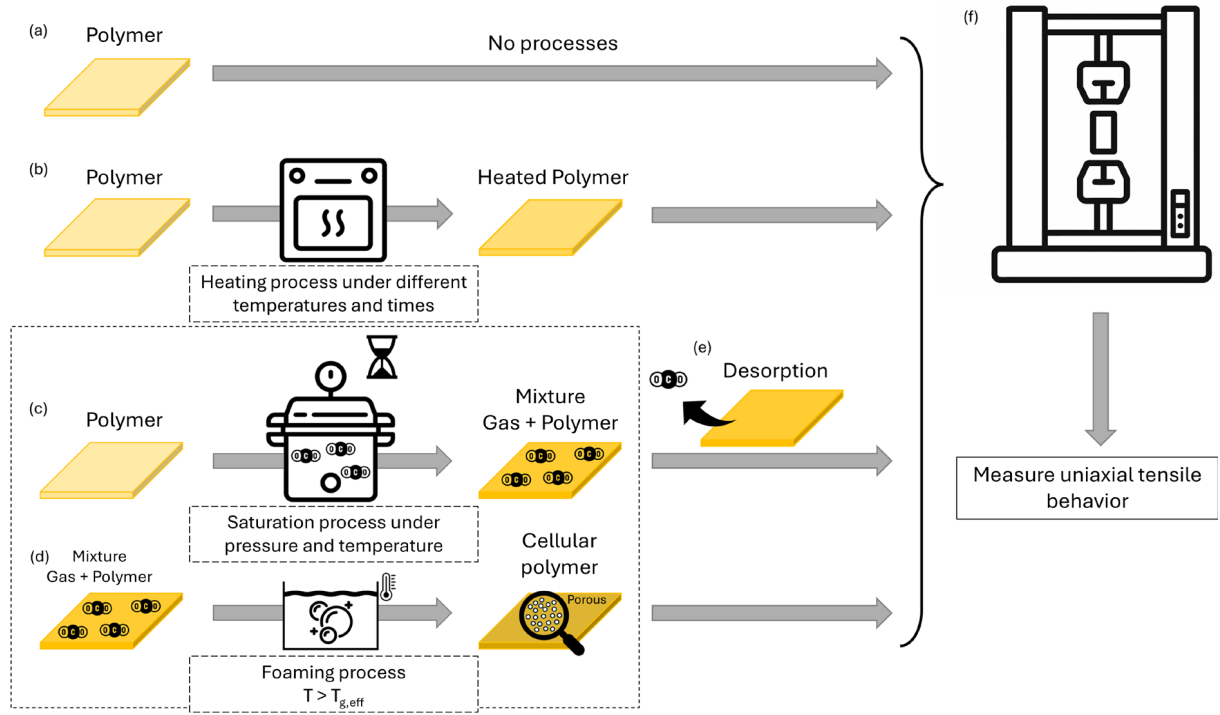

**Figure 2.** Schematic representation of the different processes applied to the samples to investigate their mechanical behavior. (a) Untreated PEI. (b) Thermal treatment at various temperatures, both below and above the $T_g$, for different times. (c) $CO_2$ saturation under controlled pressure and temperature, during which the gas diffuses into the polymer matrix. (d) Thermal foaming above the $T_{g,eff}$ of the $CO_2$-saturated polymer immersed in a high-temperature bath. Steps (c) and (d) correspond to the gas dissolution foaming route steps for producing cellular materials and can induce molecular reconfiguration in the solid phase of the polymer. (e) Desorption after saturation. (f) Uniaxial tensile tests are conducted to observe changes in mechanical behavior.

## 2.2. Effect of Thermal Treatment on the Polymer Matrix: Structural Aging of the Matrix

In the fabrication process of nanocellular polymers via gas dissolution foaming, the final stage is thermal foaming. This step involves heating the polymer above its $T_{g,eff}$, (below the reference $T_g$ in the absence of $CO_2$, which for PEI is 216 °C, but above the $T_{g,eff}$ of about 95 °C reported by Miller and co-workers under saturation conditions of 5 MPa at room temperature.[18] This effective value is known to vary with the specific saturation conditions). Since this is the last stage of processing, it is reasonable to consider that it might leave a significant structural imprint on the final material. Therefore, it is essential to evaluate how the thermal history, independently of the presence of gas, affects the mechanical properties of PEI through accelerated molecular relaxation processes.

**Figure 3** presents a systematic study of the effect of thermal treatment on untreated solid PEI, with no prior $CO_2$ exposure. Details of the procedure can be found in the *Experimental Section*.

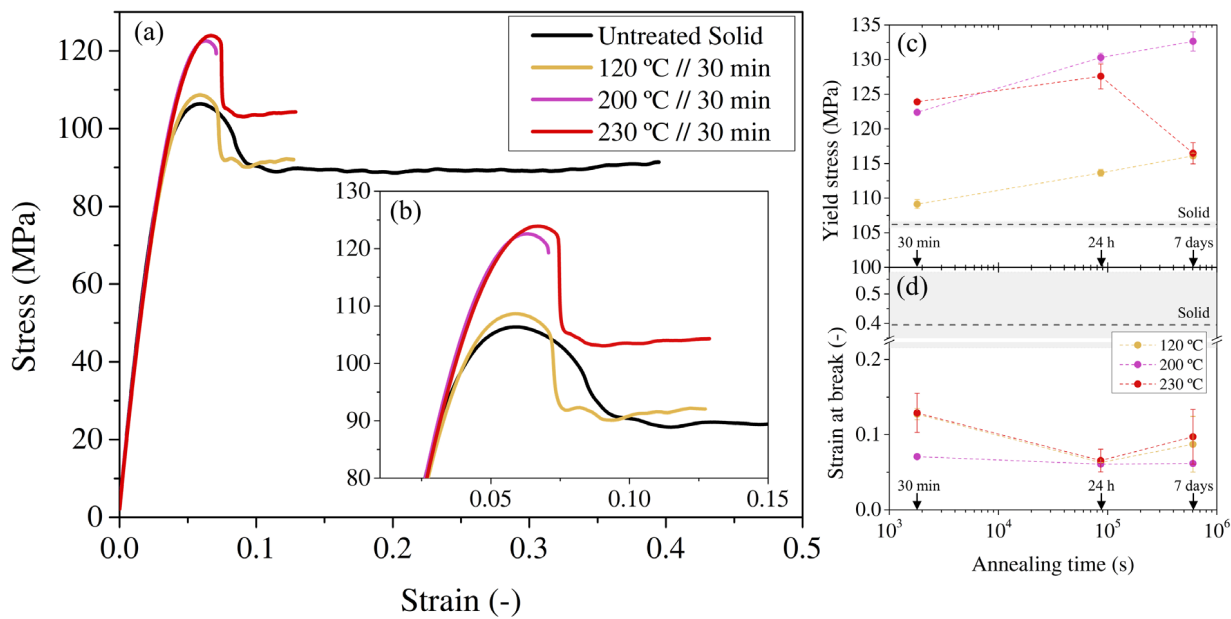

**Figure 3**. (a) Uniaxial tensile stress-strain data of untreated PEI and thermally-treated samples at different temperatures for 30 minutes. (b) Zoom of the yield region corresponding to the same previous conditions. (c) Evolution of yield stress as a function of annealing time at different temperatures. (d) Evolution of strain at break as a function of annealing time at different temperatures. The legend shown in (d) also applies to (c).

Specifically, **Figure 3a** displays the stress-strain curves of samples heated for 30 minutes at three representative temperatures: 120 °C (well below $T_g$), 200 °C (close to $T_g$, but still below), and 230 °C (above $T_g$). These specific temperatures were selected based on the fact that the $T_g$ of PEI is approximately 216 °C, as discussed in *Experimental Section*. In all cases, a notable reduction in strain at break is observed compared to the untreated solid. Additionally, the yield stress values differ significantly, which is especially relevant as this parameter is closely related to the thermomechanical history of the polymer.[24–26]

An increase in yield stress is typically associated with a denser and more compact polymer architecture, implying a reduction in the specific volume.[27] Conversely, a decrease in yield stress suggests a more open structure, with increased segmental mobility and weaker intermolecular interactions. As shown more clearly in **Figure 3b** (a zoom of the yield region), all thermal annealing treatments selected in this work lead to an increase in this parameter, suggesting progressive structural densification as temperature increases. This behavior aligns with the well-known phenomenon of physical aging, whereby exposure to sub-$T_g$ temperatures allows chains relax to a more stable state, increasing the yield stress and reducing the polymer's capacity for plastic deformation in uniaxial tension.[26,28] Even when annealing at a temperature exceeding $T_g$ (treatment at 230 °C), the polymer physically ages (corresponding to an increase of yield stress) for annealing times below or on the order of 1 day. For longer annealing times at $T$ = 230 °C, the polymer starts to thermally rejuvenate, but this process is relatively slow: after 7 days the polymer exhibits a yield stress close to that of an annealed sample at $T$ = 120 °C (see **Figure 1c**). This process is schematically illustrated in **Figure 4,** which should be regarded as a conceptual representation. While the actual molecular mechanisms are certainly more complex, this simplified view provides a useful framework for interpreting the experimental observations under the annealing time and temperatures evaluated in this work: starting from a typical amorphous structure of PEI (**Figure 4a**), thermal treatments, both below and above $T_g$, induce a reorganization towards a denser structure with reduced free volume between chains (**Figure 4a.2** and **Figure 4a.3**). This structural evolution, driven by thermal relaxation of the molecular chains, explains the progressive increase in yield stress observed experimentally.

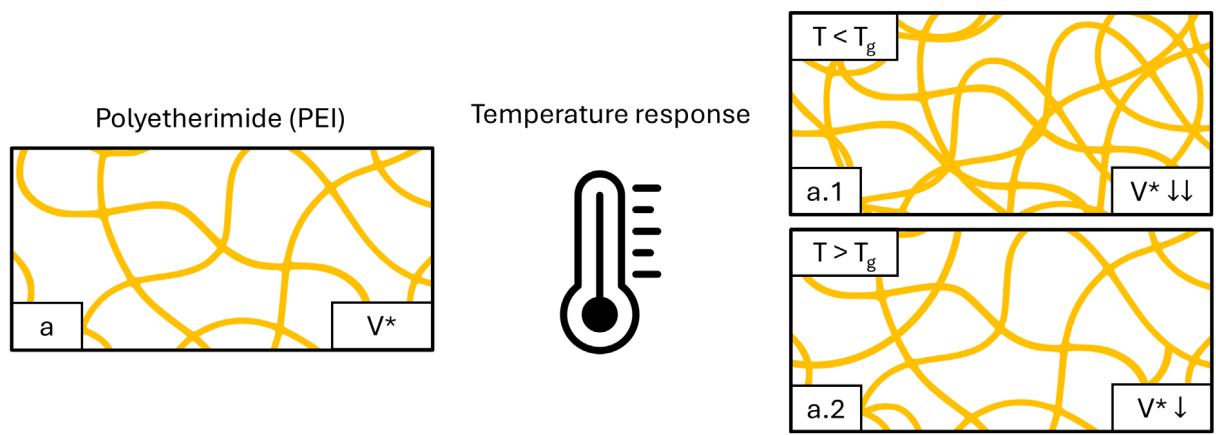

**Figure 4**. Schematic representation of the structural evolution of amorphous PEI during thermal treatment. (a) Initial amorphous configuration. (a.1) Molecular rearrangement after sub-$T_g$ annealing, leading to reduced specific volume ($V^*$). (a.2) Compact structure after annealing above $T_g$ highlighting the densification that persists after cooling. Note that the schematics reflect the structural trends observed under the specific annealing conditions applied in this work.

**Figure 3c** and **Figure 3d** further analyze the evolution of this response (yield stress and strain at break respectively) as a function of annealing time (30 minutes, 24 hours, and 7 days). For temperatures below $T_g$, a continuous increase in yield stress and a marked decrease in strain at break are observed with longer heating times. At 120 °C, the yield stress increases from 109 MPa (30 min) to 114 MPa (24 h) and 116 MPa (7 days), while the strain at break drops from 0.127 to 0.06 and 0.09, respectively. At 200 ºC, yield stress rises from 122 MPa (30 min) to 130 MPa (24 h) and 133 MPa (7 days), while strain at break decreases from 0.07 to 0.06 and remains at 0.06. This points to an accelerated aging process, in which the polymer matrix progressively relaxes towards a thermodynamic equilibrium, corresponding to an increased resistance to plastic flow (and a reduction in tensile ductility, see e.g. the work of van Melick and co-workers for a detailed discussion on this topic).[29] Above $T_g$, the yield stress seems to reach a maximum after 24 hours of annealing, increasing from 124 MPa (30 min) to 128 MPa (24 h), and then decreasing to 117 MPa after 7 days (although it does not drop below that of the original untreated solid within the explored range of annealing times). Meanwhile, the strain at break remains consistently low from 0.13 (30 min) to 0.07 (24 h) and then increases again to 0.10 at 7 days.[27]

In summary, thermal treatments carried out in this research in the absence of a prior $CO_2$ saturation process tend to increase yield stress and reduce ductility, which is in clear contrast with the observations in **Figure 1a** illustrating that nanocellular polymers exhibit a lower yield stress and a higher ductility than untreated, solid polymer when tested in uniaxial tension. Recalling that the materials were foamed at 130 ºC and 150 ºC for ultra-nano and nanocellular samples, respectively, exceeding the $T_{g,eff}$ of the saturated system (approximately 95 ºC). Thus, the present thermal treatment results are a relevant reference for comparison, as they isolate the effect of temperature under conditions similar to those experienced during foaming, but without the influence of gas. Again, this suggests that the origin of the mechanical improvements must lie elsewhere in the process. The next section focuses on the effect of $CO_2$ saturation, which is well known to act as a plasticizing step that enhances segmental mobility in the polymer matrix, even at temperatures below the reference $T_g$. For this reason, it represents a unique stage in the foaming process that can alter the polymer structure without requiring thermal activation.

**2.3. Effect of $CO_2$ Saturation on the Polymer Matrix: Evidence of Structural Rejuvenation**

After ruling out thermal treatment as the origin of the enhanced ductility observed in nanocellular polymers, since it actually induces an increase in yield stress and embrittlement, the isolated effect of $CO_2$ saturation was investigated. For this purpose, solid PEI samples were exposed to different saturation pressure and temperature conditions to reach solubility levels ranging from approximately 5% to 25% by weight. Details can be found in the *Experimental Section* (**Table 1**). After saturation, the samples were allowed to fully desorb the gas at room temperature (that means at a temperature far below the $T_{g,eff}$), thereby avoiding pore nucleation and growth and ensuring no presence of $CO_2$ in the polymer during mechanical testing (done after long desorption time, enough for the samples to completely desorb the gas). This approach enabled the evaluation of the impact of $CO_2$ on the polymer matrix without the implications of the foaming stage (void nucleation and growth) and the presence of $CO_2$ in the polymer matrix. **Figure 5** presents the results obtained from uniaxial tensile and impact tests, aimed at evaluating the mechanical response of PEI after saturation process with different amounts of $CO_2$. In **Figure 5a**, the stress-strain curves of untreated solid PEI and the various saturated samples are shown. Two main trends can be observed: (i) a moderate improvement in strain at break in comparison to the untreated solid, starting from 10% solubility (see **Figure 5d**), and (ii) a notable decrease in yield stress, occurring at lower strains and reaching significantly lower values than those of the reference solid (~106 MPa). This reduction in yield stress becomes more pronounced with increasing $CO_2$ content, dropping to 99 MPa at 10% solubility. From 15% onward, the values appear to stabilize around 95 to 96 MPa, remaining nearly constant even as solubility increases up to 25%. This trend is more clearly observed in **Figure 5b** (zoomed view of the yield region) and is represented as a plateau beyond 15% in **Figure 5c**.

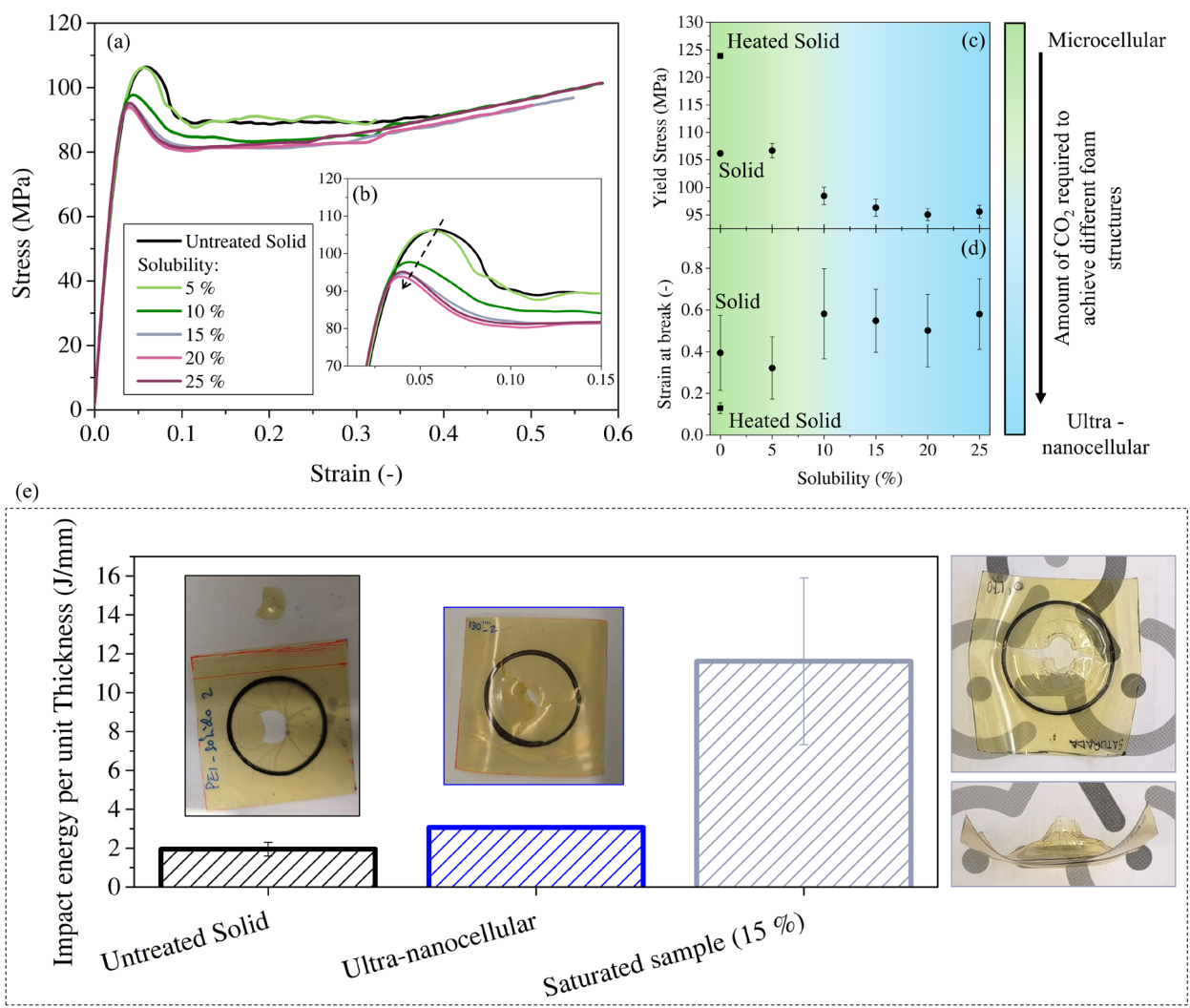

**Figure 5**. (a) Representative stress-strain data of untreated PEI and PEI saturated with different $CO_2$ solubility levels when tested in uniaxial tension (fully desorbed prior to testing). (b) Zoom view of the yield region to highlight differences in yielding behavior. (c) Yield stress as a function of $CO_2$ solubility. (d) Strain at break as a function of solubility. Datapoints associated with the *Heated Solid* at 230 °C during 30 min from **Figure 3** are included in (c) and (d) for comparison and are annotated with a square symbol. Subfigures (c) and (d) include a color scale representing the amount of gas required to generate different types of structures during the foaming process. (e) Impact test results for untreated solid PEI, an ultra-nanocellular PEI, and a sample saturated at 15% solubility, accompanied by post-fracture images illustrating the transition in fracture mode among the different materials.

In **Figure 5c** and **Figure 5d**, the points at 0% of solubility correspond to both the untreated PEI and a solid PEI sample previously heated above its $T_g$ for 30 min (square symbol). This annealed sample, which was not exposed to any $CO_2$ treatment and was already discussed in the previous section, exhibits the highest yield stress (close to 124 MPa) and lowest elongation at break (close to 0.13) among all selected annealing temperatures for a 30-minute annealing time. On the other hand, the observed decrease in yield stress and the slight increase in elongation at break in $CO_2$-saturated samples indicates that $CO_2$ effectively rejuvenates the polymer, reinforcing the hypothesis that the saturation process itself (when the $CO_2$ concentration exceeds a threshold value) modifies the mechanical behavior of the material. Elevated levels of $CO_2$ saturation erase the prior structural relaxation (physical aging) history of the polymer chains, thereby reducing the yield stress and enhancing ductility. This correlation is consistent with the well-established relationship between yield stress and strain localization: polymers with higher yield stresses typically exhibit more pronounced strain softening after yielding, which promotes strain localization and leads to brittle failure. In contrast, a reduction in yield stress mitigates this strain softening, favoring more homogeneous deformation and thus increasing ductility.[30]

From a physical standpoint, this phenomenon could be explained as shown in **Figure 6**.[31] At low solubility levels, $CO_2$ tends to occupy the polymer's free volume without substantially altering the polymer's state variables or its position in the potential energy landscape (**Figure 6a.1**), this would correspond to the scenario of PEI saturated by 5% of $CO_2$ exhibiting mechanical behavior similar to that of the solid. In this scenario, once the gas is fully desorbed, there is no significant molecular re-organization due to saturation (**Figure 6a.2**). However, above 10% solubility, excessive gas molecules irreversibly increase the average distance between the chains, thereby decreasing the intermolecular interactions between the polymer chains (**Figure 6a.3**). This reconfiguration would also correspond to an increased specific free volume, as shown in **Figure 6a.4** and leads to a change in macroscopic deformation and failure behavior as shown in **Figure 5** (decrease in yield stress and an increase in tensile ductility).

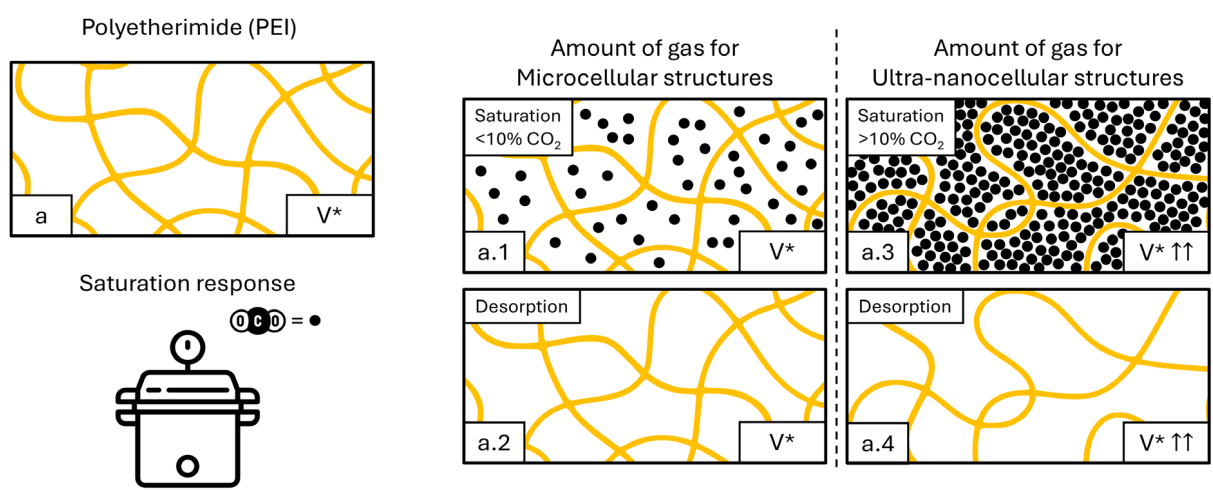

**Figure 6**. Schematic evolution of PEI molecular architecture upon $CO_2$ saturation. (a) Initial amorphous structure. (a.1) After saturation at low $CO_2$ content, the gas fills the free volume with (a.2) minimal structural change. (a.3) At high $CO_2$ content, gas molecules separate polymer chains and dilute the polymer. (a.4) Final configuration retained after complete gas desorption.

Notably, the onset of this effect aligns with the solubility threshold (>10%) above which nanocellular structures have been reported in nanocellular PEI materials as indicated by the blue region in the color scale presented in **Figure 5c** and **Figure 5d**. This suggests that the macromolecular interactions induced by gas saturation play a role in the enhanced properties and transitions observed in cellular polymers, even before any void is formed during the subsequent foaming step.

Finally, **Figure 5a** shows that towards the end of the tensile test, all curves tend to converge prior to fracture which is a well-known observation associated with mechanical rejuvenation.[26,32] This observation also supports the idea that $CO_2$ simply acts as a rejuvenator of the polymer system. To further investigate the effects of gas saturation on mechanical performance for more complicated loading states, impact tests were performed (**Figure 5e**). In this case, the sample saturated with 15% of $CO_2$ exhibited not only a significantly higher energy absorption than the solid, but also a clear transition from brittle to ductile fracture, as evidenced by the fracture morphology: while the solid sample fails through multiple cracks, the saturated sample undergoes significant plastic deformation, forming a well-defined necking region, similar to that observed in the ultra-nanocellular material. These findings confirm that $CO_2$ saturation, even in the absence of foaming, induces modifications to the internal architecture of PEI, which are preserved after complete gas desorption. This reveals a previously overlooked mechanism that could help to explain the mechanical transitions observed in nanocellular polymers. However, it should be noted that these effects may diminish over time or under thermal exposure, as physical aging of the polymer in the cellular structure may gradually result in recovery of the untreated, solid polymer's constitutive behavior, a point that deserves further investigation. The first steps of this hypothesis are taken in the following section.

## 2.4. Thermal stability of the saturation effect: role of β-relaxation and its implications for foaming

Once it has been demonstrated that $CO_2$ saturation induces a modification in the internal architecture of PEI, evidenced by the decrease in yield stress, it becomes essential to assess whether this structural configuration is thermally stable. This analysis is particularly relevant considering that, after saturation, there is a foaming step during which the material is exposed to elevated temperatures above the $T_{g,eff}$ of the saturated system, which could potentially embrittle the polymer due to accelerated chain relaxation processes (accelerated physical aging). To address this question, **Figure 7** compiles mechanical and thermo-mechanical data of saturated PEI samples subjected to short thermal treatments at different temperatures. Through the combination of uniaxial tensile analysis and DMA, it becomes possible to evaluate whether the structural changes induced by saturation are preserved or reversed upon heating.

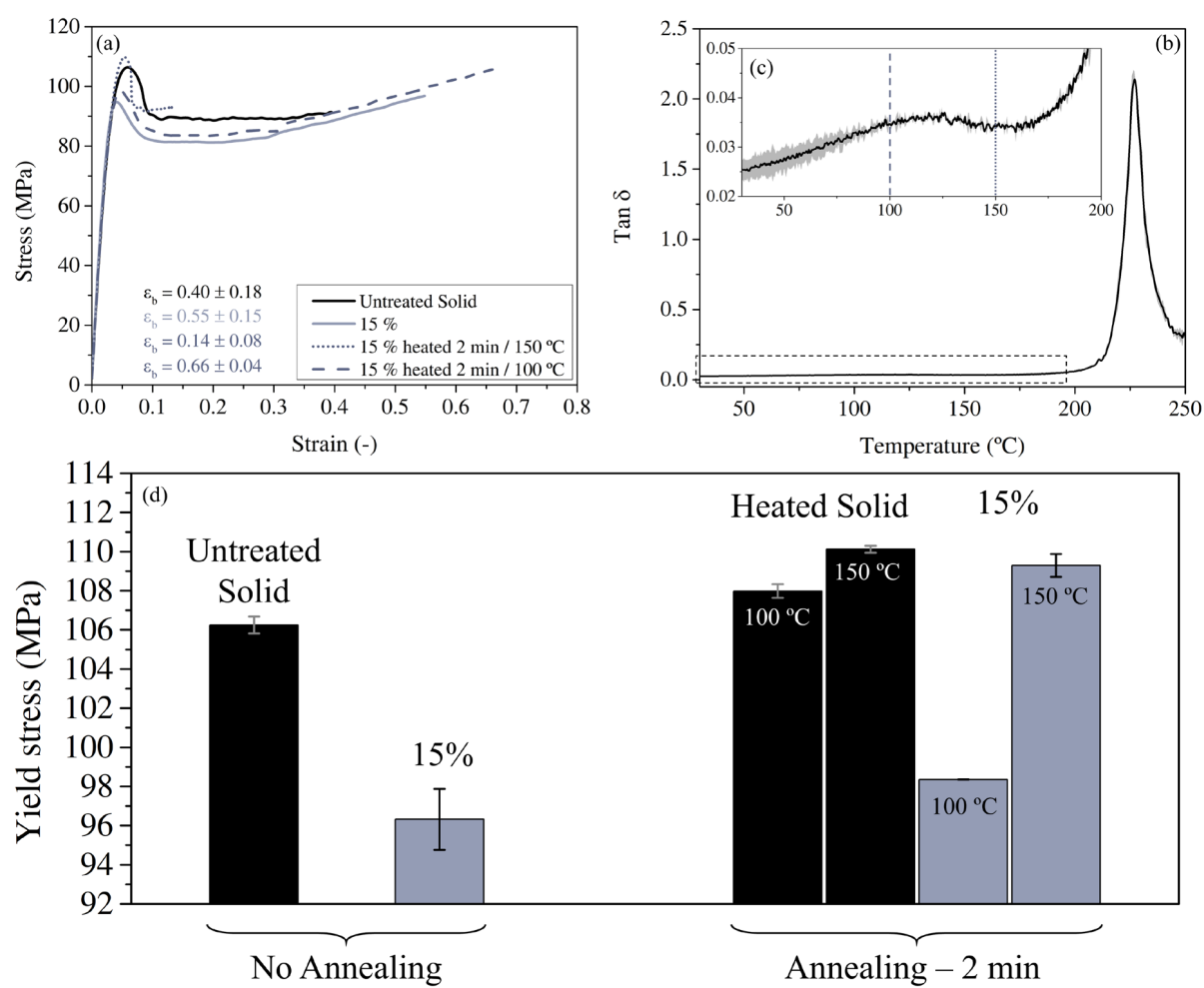

**Figure 7**. (a) Stress-strain data of untreated PEI samples and samples saturated at 15% $CO_2$ when testes in uniaxial tension, with and without subsequent thermal treatment (2 minutes at 100 °C and 150 °C). The strain at break is reported for each case, including the corresponding standard deviation, following the same order from top to bottom as appears in the legend. (b) *tan δ* curve obtained by DMA of untreated solid PEI. (c) Zoomed view of the *tan δ* curve in the 30 - 200 °C temperature range. (d) Yield stress as a function of thermal treatment for untreated solid and 15% saturated samples, before and after 2 minutes of heating at 100 °C and 150 °C.

**Figure 7a** shows the stress-strain curves for untreated solid PEI, a sample saturated at 15% solubility, and two additional samples also saturated at 15% but subsequently heated for 2 minutes at 100 °C and 150 °C, respectively, simulating a foaming step without $CO_2$ into the polymer matrix. The results reveal that the sample heated at 100 °C maintains a mechanical response similar to that of the unheated saturated one (yield stress close to 98 MPa). In contrast, the sample heated at 150 °C exhibits a significantly higher yield stress (close to 109 MPa), even exceeding that of the original solid, suggesting a structural reorganization towards stronger intermolecular interactions. These results align with those observed in the thermal treated samples (**Figure 3**). Thus, these findings indicate that the rejuvenation effect by $CO_2$ can be thermally reversed if a critical combination of foaming temperature and time is surpassed.

To interpret this behavior, the loss tangent (*tan δ*) of the untreated PEI, measured through DMA, as explained in *Experimental Section*, is presented in **Figure 7b**. As expected, the $T_g$ of PEI appears above 200 °C. However, a closer inspection of the temperature range between 30 °C and 200 °C (**Figure 7c**) reveals a second *tan δ* peak between 100 °C and 150 °C, corresponding to a β-relaxation. This transition is associated with local segmental motions of chains or side groups of the macromolecular chains.[33,34] Therefore, heating above this threshold activates enough mobility to allow structural relaxation of the chains, effectively erasing the rejuvenation effect of saturation.

**Figure 7d** reinforces this hypothesis by showing the evolution of yield stress as a function of annealing time for different temperatures. At 100 °C, both the solid and saturated PEI samples maintain similar increases in yield stress around 2% with clearly different yield stress values. However, after treatment at 150 °C, the two materials converge toward a common yield stress value approximately equal to 110 MPa. This convergence suggests that once β-relaxations are activated, both systems tend towards similar positions in the polymer system's potential energy landscape.

This phenomenon could also explain why the effect of $CO_2$ on the properties of the solid polymer has not been reported in more widely studied polymers such as PMMA or polycarbonate (PC). In those materials, β-relaxation occurs at much lower temperatures, close to or even below room temperature, so the effect of $CO_2$ might dissipate rapidly after desorption, without leaving a persistent structural footprint. [35,36]

Based on the results of **Figure 7**, one could expect that the effect of elevated temperature during foaming might erase the effect of prior $CO_2$ presence. In contrast, in the case of PEI, this effect can persist even after the foaming process, resulting in an increased ductility as proven in previous studies like presented in **Figure 1c** and mechanical behavior shown in **Figure 5**. The complete hypothesis that could explain this phenomenon is illustrated in **Figure 8**. Starting from a conventional amorphous state (**Figure 8a**), the material is first saturated with $CO_2$ until reaching a solubility level sufficient to produce a nanocellular structure (**Figure 8a.1**), as described in the previous section. After depressurization, the foaming step takes place. During this stage, $CO_2$ has not yet been fully desorbed and remains present within the polymer matrix (**Figure 8a.2**). It has been demonstrated that significant amounts of gas can remain trapped in the solid phase of PEI even after prolonged desorption times.[17] Similar behavior has also been reported in PMMA-based foams, where $CO_2$ retention was detected even after the foaming step.[37,38] This residual gas maintains and prevents thermal relaxation of the structure. It is only after complete desorption of the gas, once the nanocellular polymer has stabilized at room temperature, that the polymer structure consolidates its final state. At that point, the solid phase of the foamed material inherits the mechanical properties previously observed in the non-foamed, $CO_2$-saturated PEI, now embedded in a stabilized porous architecture.

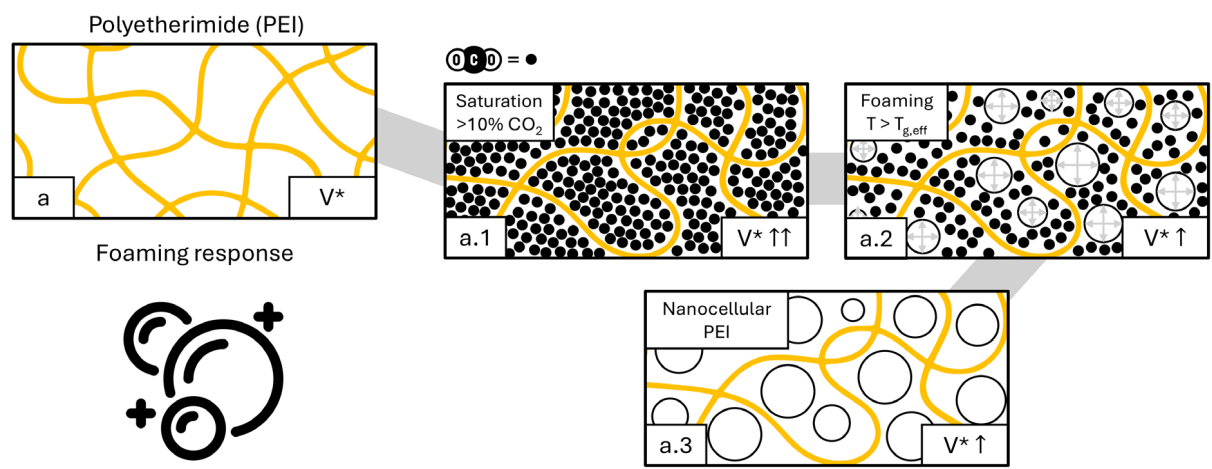

**Figure 8**. Schematic description of the structural evolution of PEI during and after foaming. (a) Initial amorphous state. (a.1) $CO_2$ saturation increases the specific volume of the polymer; this is associated with erasing prior structural relaxation of the chains. (a.2) During foaming, gas remains within the matrix, preventing chain relaxation. (a.3) After complete desorption at room temperature, the $CO_2$-induced structural changes are preserved and consolidated.

## 3. Conclusion

This work demonstrates that the change in mechanical behavior observed in nanocellular PEI cannot be explained solely by the presence of the nanoporous structure. While literature has traditionally linked improvements in toughness, impact resistance, and ductility to the stress-relieving effect of nanopores, these results reveal that the rejuvenated polymeric matrix induced by $CO_2$ saturation plays a role alongside nanopores.

Beyond a solubility threshold close to 10%, a decrease in yield stress appears after complete gas desorption, revealing that $CO_2$ saturation may act as a structural rejuvenation mechanism (erasing prior chain relaxation history). This effect was confirmed by both uniaxial tensile and impact tests, in which non-foamed saturated samples (without $CO_2$ inside) exhibited a mechanical response even superior to that of ultra-nanocellular polymers, which already outperformed the untreated solid PEI.

However, this structural state is not entirely stable under thermal exposure. A short (2 min) thermal treatment at 150 °C is sufficient to remove the $CO_2$-induced rejuvenation effect, restoring the mechanical properties of the saturated PEI to values nearly identical to those of the untreated solid. This reversion is attributed to the activation of the material's β-relaxation, as revealed by DMA. In contrast, in the case of PEI, the saturation effect can persist even after foaming. During the foaming stage, $CO_2$ is not yet fully desorbed and remains within the polymer matrix. This presence temporarily limits chain mobility, preventing thermal relaxation of the chains. Only after complete gas desorption, when the nanocellular polymer has stabilized at room temperature, does the polymer architecture consolidate its final state. At this point, the solid phase of the nanocellular polymer inherits the mechanical properties previously observed in the non-foamed saturated PEI. However, once the nanocellular polymer architecture has consolidated into its final state after complete gas desorption, continued thermal exposure (e.g., through subsequent annealing or prolonged use at elevated temperatures in an application) could accelerate the physical aging of the material. This process could induce a recovery of yield stress and an increase in strain softening, reversing the initial rejuvenation effect and leading to a transition to brittle behavior. This brittle behavior of the parent polymer material in the nanocellular material would correspond to the behavior of the 'as-received' solid , which has been subjected to thermal rejuvenation during its initial processing, but an exhibiting brittle behavior over time due to physical aging. It should be emphasized that this is a hypothesis, as the large gap between the $T_g$ of PEI and room temperature suggests that very long timescales may be required for the material to recover the structural state of the as-received solid. Further investigation will be needed to confirm this.

Altogether, these findings reveal that the mechanical performance of nanocellular PEI materials cannot be fully understood without considering the direct impact of rejuvenation of the polymer during the saturation step. Recognizing that the behavior of these materials results from a synergy between the porous architecture and a transformed polymer matrix is essential for guiding the design of next-generation high-performance functional cellular polymers.

## 4. Experimental Section

*Materials:* The polymer employed in this study was ULTEM™ Resin 1000, a commercial grade of PEI. It was supplied in sheet form with a thickness of 0.17 mm. The material has a density of 1.28 g·cm$^{-3}$ and its $T_g$, determined by DSC, is approximately 216 °C. The $CO_2$ used in saturation and foaming processes was medical grade, with a purity above 99.9 %.

*Samples preparation:* PEI samples were prepared in different formats depending on the type of mechanical test to be conducted. For tensile testing, rectangular specimens of 8 × 2 cm$^2$ were cut from PEI sheets. After undergoing the corresponding treatment (heating, saturation, or foaming), these specimens were machined into dog-bone shapes in accordance with ISO 527-2 (specimens' type 1BA). For impact tests, the samples were cut into square shapes of 6 × 6 cm$^2$. To provide a clear overview of the experimental conditions and treatments applied, **Table 1** summarizes all materials investigated in this study, organized according to sample type, saturation and/or thermal treatment conditions, and the applied foaming process.

**Table 1**. Summary of the treatments and experimental conditions applied to the PEI samples. When a parameter is reported with two values, it indicates that both were used independently in different sample batches, rather than sequentially applied to the same specimens.

| Sample | $CO_2$ Saturation ($P_{sat}$, $T_{sat}$, Sol) | Thermal treatment (T, t) | Foaming ($T_f$, $t_f$, $\phi$, $\rho_r$) |
|---|---|---|---|
| Untreated solid | - | - | - |
| Annealing sub-$T_g$ | - | 120 °C, 200 °C <br> 30 min, 24 h, 7 days | - |
| Annealing supra-$T_g$ | - | 230 °C <br> 30 min, 24 h, 7 days | - |
| Non-foamed saturated sample | 1 MPa, 24 °C <br> ~ 5% | - | - |
| Non-foamed saturated sample | 6 MPa, 50 °C <br> ~ 10% | - | - |
| Non-foamed saturated sample | 6 MPa, 24 °C <br> ~ 15% | - | - |
| Non-foamed saturated sample | 50 MPa, 24 ºC ~ 20% | - | - |
| Non-foamed saturated sample | 50 MPa, 0 ºC ~ 25% | - | - |
| Non-foamed saturated sample + Thermal treatment without $CO_2$ | 6 MPa, 24 ºC ~ 15% | 100 ºC, 150 ºC 2 min | - |
| Nanocellular polymer | 6 MPa, 50 ºC ~ 10% | - | 150 ºC, 30 s, 780 nm, 0.82 |
| Ultra-nanocellular polymer | 6 MPa, 24 ºC ~ 15% | - | 130 ºC, 30 s, 35 nm, 0.82 |

*Annealing process:* To independently evaluate the effect of heat on the polymer matrix, without the influence of gas or porosity, selected samples were subjected to thermal treatments (annealing) show in **Table 1**. This process involved placing the samples in an oven at a controlled temperature for 30 minutes, 24 hours, and 7 days, using three different conditions: 120 ºC and 200 ºC (both below $T_g$ of PEI), and 230 ºC (above $T_g$).

*Gas dissolution set-up:* For the remaining samples, saturation and foaming were carried out using the gas dissolution foaming technique. During the saturation step, the samples were exposed to $CO_2$ at various saturation pressures and temperatures conditions for 24 hours. This step was performed in a high-pressure autoclave (High Pressure Chemical Reactor, 100 ml, Supercritical Fluid Technologies Inc., Newark, DE, USA), equipped with a pressure pump (SFT-10 model, same supplier), a temperature controller (Polyscience AD15R-40-A11B 15L), and a thermal jacket to ensure stable saturation conditions.

*Saturation process:* In the case of non-foamed saturated samples, intended to isolate and study the effect of gas on the polymer matrix, five different saturation conditions were used to achieve a solubility range from approximately 5% to 25%, in steps of about 5%. The selected conditions, shown in **Table 1**, after removal from the autoclave, the samples were fully desorbed at room temperature. Some of these saturated samples were also subjected to post-saturation thermal treatment in order to study the combined effect of $CO_2$ and heat on the polymer structure.

*Foaming process:* For cellular structure generation, two saturation temperatures were used as can be seen in **Table 1**, while keeping pressure constant at 6 MPa, depending on the desired cellular structure: 24 ºC for ultra-nanocellular polymers (transparent samples) and 50 ºC for nanocellular polymers (opaque samples). Foaming was carried out after 40 seconds of desorption by immersing the samples in a silicone oil bath for 30 seconds, at temperatures high enough to exceed the $T_{g,eff}$ of the saturated system, thus promoting cell growth. Samples saturated at 24 ºC were foamed at 130 ºC, while those saturated at 50 ºC were foamed at 150 ºC, in both cases adjusting conditions to obtain distinct cellular structures with comparable densities.

*Characterization:* The solubility of $CO_2$ absorbed by the samples was determined from the mass increase observed after the saturation process, which serves as an indicator of gas solubility. Since depressurization of the autoclave leads to partial gas loss before weighing, an indirect approach was employed by extrapolating the desorption curve (mass vs. time) back to zero time following Fick's laws.[39] Measurements were performed using a Mettler Toledo AT261 balance connected to a digital interface, allowing continuous monitoring of mass loss during the first minutes after extraction. The zero-time mass was obtained by fitting the linear portion of the desorption data between 70 and 140 seconds, where the time evolution follows a square root dependency. This method, thoroughly validated in previous works on gas solubility and diffusion in polymers, has proven to be a reliable tool for estimating the amount of gas absorbed under equilibrium conditions.[40]

The tensile mechanical properties of the samples were evaluated at a controlled temperature of 23 ± 2 ºC and a relative humidity of 50 ± 10%, using an Instron universal testing machine (model 5500R6025) and a Zwick/Roell testing machine. Tests were carried out in accordance with ISO 527-2, applying a strain rate of 10 mm/min (corresponding to a global strain rate equal to $2.9 \times 10^{-3}$ s$^{-1}$).

The impact properties of the samples were measured using an instrumented falling weight impact (IFWI). Tests were conducted at the same temperature and humidity conditions as tensile tests, using a custom-built device developed by CellMat Laboratory and manufactured by Microtest.[41] The system employed a 12.7 mm diameter hemispherical striker equipped with a KISTLER type 9333A force sensor mounted on top, operating at a data acquisition frequency of 55.3 kHz. Striker displacement was measured with an LDS90/40 laser triangulation sensor (LMI Sensors-95), enabling accurate monitoring of the dynamic response of the material. The incident energy was 55.66 J, calculated from a striker mass of 5.75 kg and an impact velocity of 4.4 m/s. All tests and data analysis were carried out according to ISO 6603/1-2 standards.

The loss factor (*tan δ*) was measured using a DMA 8000 (PerkinElmer) in tensile mode with a frequency of 1 Hz and a heating rate of 5 ºC·min$^{-1}$. The temperature sweep was performed from 30 ºC to 250 ºC, covering both the β-relaxation and the glass transition region.

**Acknowledgements**

The authors would like to thank Laura Izaguirre Altuna for some tensile and impact trials, Blanca Calvo Cabezón for DMA measurements and M. Isabel Muñoz de Frutos for some heating experiments. Lizalde-Arroyo is also deeply grateful to the P&P group at TU/e, for their warm welcome, kind hospitality and valuable time shared during the research stay.

**Data Availability Statement**

The data that supports the findings of this study are available from the corresponding author, upon reasonable request.

**References**

1. Costeux S. CO 2 -blown nanocellular foams. *J Appl Polym Sci*. 2014;131(23). doi:10.1002/app.41293

2. Rodríguez Pérez MA, Martín de León J, Bernardo García V. *Nanocellular Polymers*. De Gruyter; 2023. doi:10.1515/9783110756135

3. Li X, Zong Y, Li W, et al. Achieving Ultrahigh Transparency and Superior Mechanical Properties in Flexible Polyimide Nanofoams Through CO 2 Foaming for Thermal Insulation. *Adv Funct Mater*. 2024;34(49):1-11. doi:10.1002/adfm.202409498

4. Miller D, Kumar V. Microcellular and nanocellular solid-state polyetherimide (PEI) foams using sub-critical carbon dioxide II. Tensile and impact properties. *Polymer (Guildf)*. 2011;52(13):2910-2919. doi:10.1016/j.polymer.2011.04.049

5. Gebremedhin KF, Tolcha SD, Yeh SK. Fabrication of flat and sizeable nanocellular polymethyl methacrylate (PMMA) foam with tunable thermal conductivity. *Polym Eng Sci*. 2024;(July):4973-4992. doi:10.1002/pen.26895

6. Sánchez-Calderón I, Bernardo V, Martín-de-León J, Rodríguez-Pérez MÁ. Thermal conductivity of low-density micro-and nanocellular poly(methyl-methacrylate) (PMMA): Experimental and modeling. *Mater Des*. 2022;221. doi:10.1016/j.matdes.2022.110938

7. Liu S, Duvigneau J, Vancso GJ. Nanocellular polymer foams as promising high performance thermal insulation materials. *Eur Polym J*. 2015;65:33-45. doi:10.1016/j.eurpolymj.2015.01.039

8. Torre J, Cuadra-Rodríguez D, Barroso-Solares S, et al. Understanding molecular confinement in polymeric nanoporous materials via infrared spectroscopic measurements. *React Funct Polym*. 2025;216(March). doi:10.1016/j.reactfunctpolym.2025.106401

9. Martín-de León J, Van Loock F, Bernardo V, Fleck NA, Rodríguez-Pérez MÁ. The influence of cell size on the mechanical properties of nanocellular PMMA. *Polymer (Guildf)*. 2019;181(June):121805. doi:10.1016/j.polymer.2019.121805

10. Martín-de León J, Pura JL, Bernardo V, Rodríguez-Pérez MÁ. Transparent nanocellular PMMA: Characterization and modeling of the optical properties. *Polymer (Guildf)*. 2019;170(February):16-23. doi:10.1016/j.polymer.2019.03.010

11. Lizalde-Arroyo F, Bernardo V, Rodríguez-Pérez MÁ, Martín-de León J. Exploring transparent ultra – nanocellular high-performance polymers: Polyetherimide (PEI). *Mater Des*. 2025;255(June):114222. doi:10.1016/j.matdes.2025.114222

12. Sorrentino L, Aurilia M, Iannace S. Polymeric foams from high-performance thermoplastics. *Adv Polym Technol*. 2011;30(3):234-243. doi:10.1002/adv.20219

13. Holl MR, Kumar V, Garbini JL, Murray WR. Cell nucleation in solid-state polymeric foams: evidence of a triaxial tensile failure mechanism. *J Mater Sci*. 1999;34(3):637-644. doi:10.1023/A:1004527603363

14. Aher B, Olson NM, Kumar V. Production of bulk solid-state PEI nanofoams using supercritical CO 2. *J Mater Res*. 2013;28(17):2366-2373. doi:10.1557/jmr.2013.108

15. Verreck G, Decorte A, Li H, et al. The effect of pressurized carbon dioxide as a plasticizer and foaming agent on the hot melt extrusion process and extrudate properties of pharmaceutical polymers. *J Supercrit Fluids*. 2006;38(3):383-391. doi:10.1016/j.supflu.2005.11.022

16. Guo H, Nicolae A, Kumar V. Fabrication of high temperature polyphenylsulfone nanofoams using high pressure liquid carbon dioxide. *Cell Polym*. 2016;35(3):119-142. doi:10.1177/026248931603500302

17. Patel ZS, Sridhar S, Nadella K, Kumar V, Nuhnen A, Janiak C. Nanoporous structure development in PEI thin film foams. *J Mater Sci*. 2023;58(44):17113-17125. doi:10.1007/s10853-023-09080-4

18. Miller D, Chatchaisucha P, Kumar V. Microcellular and nanocellular solid-state polyetherimide (PEI) foams using sub-critical carbon dioxide I. Processing and structure. *Polymer (Guildf)*. 2009;50(23):5576-5584. doi:10.1016/j.polymer.2009.09.020

19. Martín-de León J, Pura JL, Rodríguez-Méndez ML, Rodríguez-Pérez MA. Viscoelastic property enhancement of polymethylmethacrylate molecularly confined within 3D nanostructures. *Eur Polym J*. 2024;214(May):113181. doi:10.1016/j.eurpolymj.2024.113181

20. Dwiwedi K, Sridhar S, Kumar V, Salviato M, Meza L. Breaking Down the Exceptional Size-Dependent Toughness of Nanocellular Foams. Published online 2025. doi:10.2139/ssrn.5208612

21. Lin Y, Tan Y, Qiu B, Shangguan Y, Harkin-Jones E, Zheng Q. Influence of annealing on chain entanglement and molecular dynamics in weak dynamic asymmetry polymer blends. *J Phys Chem B*. 2013;117(2):697-705. doi:10.1021/jp3098507

22. Pan Y, Pan X, Wang D, et al. The role of annealing on the properties and microstructure of amorphous/crystalline PPESK/PEKK system. *Polymer (Guildf)*. 2025;317(November 2024):127928. doi:10.1016/j.polymer.2024.127928

23. Mohd Yasin SB, Terry JS, Taylor AC. Fracture and mechanical properties of an impact

toughened polypropylene composite: modification for automotive dashboard-airbag application. *RSC Adv*. 2023;13(39):27461-27475. doi:10.1039/d3ra04151d

24. Hutchinson JM. Physical aging of polymers. *Prog Polym Sci*. 1995;20(4):703-760. doi:10.1016/0079-6700(94)00001-I

25. Utz M, Debenedetti PG, Stillinger FH. Atomistic Simulation of Aging and Rejuvenation in Glasses. *Phys Rev Lett*. 2000;84(7):1471-1474. doi:10.1103/PhysRevLett.84.1471

26. Meijer HEH, Govaert LE. Mechanical performance of polymer systems: The relation between structure and properties. *Prog Polym Sci*. 2005;30(8-9):915-938. doi:10.1016/j.progpolymsci.2005.06.009

27. Huang H, Huang L, Lin F, et al. Room temperature brittle-to-ductile transition in polystyrene induced by extrusion casting melt stretching: Contribution of free volume and chain entanglement. *Polymer (Guildf)*. 2024;294(November 2023). doi:10.1016/j.polymer.2024.126745

28. De Noni L, Zhu X, Andena L, et al. Effect of physical aging on scratch behavior of polycarbonate. *J Appl Polym Sci*. 2024;141(43):1-14. doi:10.1002/app.56124

29. van Melick HGH, Govaert LE, Raas B, Nauta WJ, Meijer HEH. Kinetics of ageing and re-embrittlement of mechanically rejuvenated polystyrene. *Polymer (Guildf)*. 2003;44(4):1171-1179. doi:10.1016/S0032-3861(02)00863-7

30. van Melick HGH, Govaert LE, Meijer HEH. Localisation phenomena in glassy polymers: influence of thermal and mechanical history. *Polymer (Guildf)*. 2003;44(12):3579-3591. doi:10.1016/S0032-3861(03)00089-2

31. Torres A, Soto C, Carmona J, et al. Gas Permeability through Polyimides: Unraveling the Influence of Free Volume, Intersegmental Distance and Glass Transition Temperature. *Polymers (Basel)*. 2023;16(1):13. doi:10.3390/polym16010013

32. Trejo Carrillo D, Díaz Díaz A. Nonlinear Response of a Polycarbonate in Post-Yield Cyclic Tests. *Polymers (Basel)*. 2025;17(11):1535. doi:10.3390/polym17111535

33. Sanchis MJ, Díaz-Calleja R, Jaïmes C, et al. A relaxational and conductive study on two poly(ether imide)s. *Polym Int*. 2004;53(9):1368-1377. doi:10.1002/pi.1544

34. Zhu L, Zhang Y, Xu W, et al. Crosslinked polyetherimide nanocomposites with superior energy storage achieved via trace Al2O3 nanoparticles. *Compos Sci Technol*. 2022;223(November 2021). doi:10.1016/j.compscitech.2022.109421

35. Ionita D, Cristea M, Banabic D. Viscoelastic behavior of PMMA in relation to deformation mode. *J Therm Anal Calorim*. 2015;120(3):1775-1783. doi:10.1007/s10973-015-4558-4

36. Song P, Trivedi AR, Siviour CR. Mechanical response of four polycarbonates at a wide range of strain rates and temperatures. *Polym Test*. 2023;121(February):107986. doi:10.1016/j.polymertesting.2023.107986

37. Pinto J, Reglero-Ruiz JA, Dumon M, Rodriguez-Perez MA. Temperature influence and $CO_2$ transport in foaming processes of poly(methyl methacrylate)-block copolymer nanocellular and microcellular foams. *J Supercrit Fluids*. 2014;94:198-205. doi:10.1016/j.supflu.2014.07.021

38. Van Loock F, Bernardo V, Rodríguez Pérez MA, Fleck NA. The mechanics of solid-state nanofoaming. *Proc R Soc A Math Phys Eng Sci*. 2019;475(2230):20190339. doi:10.1098/rspa.2019.0339

39. Crank J. *The Mathematics of Diffusion.*; 1975.

40. Guo H, Kumar V. Some thermodynamic and kinetic low-temperature properties of the PC-CO2 system and morphological characteristics of solid-state PC nanofoams produced with liquid CO2. *Polymer (Guildf)*. 2015;56:46-56. doi:10.1016/j.polymer.2014.09.061

41. Ruiz-Herrero JL, Rodríguez-Pérez MA, De Saja JA. Design and construction of an instrumented falling weight impact tester to characterise polymer-based foams. *Polym Test*. 2005;24(5):641-647. doi:10.1016/j.polymertesting.2005.03.011